\documentclass[12pt]{article}%
\usepackage{amsmath,amssymb,amsthm,amsfonts}
\usepackage{wasysym}
\usepackage{graphicx}
\usepackage[dvipsnames]{xcolor}
\usepackage{stackengine}

\usepackage[colorlinks]{hyperref}
\usepackage{tikz}
\usepackage[export]{adjustbox}
\usepackage{braket}

\usepackage{appendix}

\renewcommand{\epsilon}{\varepsilon}

\newcommand{\del}{\nabla}

\definecolor{red}{rgb}{0.8500, 0.3250, 0.0980}
\definecolor{green}{rgb}{0.4660, 0.6740, 0.1880}
\definecolor{yellow}{rgb}{0.9290, 0.6940, 0.1250}
\definecolor{blue}{rgb}{0, 0.4470, 0.7410}

\title{Towards a Deterministic Interpretation of Quantum Mechanics: Insights from Dynamical Systems}

\author{Aminur Rahman \thanks{ E-mail:  arahman2@uw.edu} \thanks{Department of Applied Mathematics, University of Washington, Seattle, WA} \thanks{Artificial Intelligence Institute in Dynamic Systems, University of Washington, Seattle, WA}}

\date{}

\begin{document} 

\maketitle


\begin{abstract}
Experiments violating Bell's inequality appear to indicate deterministic models do not correspond to a realistic theory of quantum mechanics. The theory of pilot waves seemingly overcomes this hurdle via nonlocality and statistical dependence, however it necessitates the existence of ``ghost waves''. This manuscript develops a deterministic dynamical system with local interactions. The aggregate behavior of the trajectories are reminiscent of a quantum particle evolving under the Schr\"{o}dinger equation and reminiscent of Feynman's path integral interpretation in three canonical examples:  motion in free space, double slit diffraction, and superluminal barrier traversal.  Moreover, the system bifurcates into various dynamical regimes including a classical limit.  These results illustrate a deterministic alternative to probabilistic interpretations and aims to shed light on the transition from quantum to classical mechanics.
\end{abstract}

\section{Introduction}\label{Sec: Intro}

In recent decades there has been both a resurgence of and fervor against deterministic theories of quantum mechanics such as pilot wave theory \cite{BohmVigier, AdvancesPilotWave} and superdeterminism \cite{DonadiHossenfelder}.  A consequence of the pilot wave theory of Bohm are ``ghost waves'' and superdeterminsm does not stand on a firm mathematical foundation, however the Copenhagen interpretation saddles us with the unrealistic collapse of the wave function.  Indeed the search for a deterministic theory is nothing new.  Pilot waves were initially proposed by de Broglie \cite{deBrogliePilotWave1927}, and more recently other deterministic theories based on relaxing measurement independence, such as that of Hall \cite{HallDeterministic2010}, have shown considerable promise.

%%%A collapse of the wave-function
Partial differential equations (PDEs) are useful in modeling the aggregate behavior of many particles acting in unison by assuming the amount of particles is so plentiful that they form a continuum.  It is no wonder that they are used to model processes like heat transfer and the propagation of waves.  One can also see the appeal in modeling the evolution of possible quantum states of a particle in this way, and using a wave-function collapse as a mathematical trick to justify a single measurement.  As a thought experiment let us consider the distribution of pollen particles in water governed by the diffusion equation
\begin{equation}
\frac{\partial u}{\partial t} = \del\cdot\left(D\del u\right);\; u(t = 0) = f(x,y),
\label{Eq: Heat}
\end{equation}
Instead of interpreting $u(x, y; t)$ as the concentration of pollen particles, one could normalize and interpret it as the probability of finding a pollen particle at a particular location $(x, y)$.  Now if we repeat the experiment with a single pollen particle we would have the same probability distribution.  If we were to make a measurement at a time $t = t_*$ and find the particle at a particular location we could interpret it as the ``diffusion function'' collapsing even though the dynamics of a single pollen particle is completely deterministic.

One of the many appeals of the Schr\"{o}dinger equation \cite{Schrodinger},
\begin{equation}
i\frac{\partial}{\partial t}\ket{\Psi} = \hat{H}\ket{\Psi}
\label{Eq: Schrodinger}
\end{equation}
is in the complex dynamics that it can describe in a simple compact form.  As shown by May \cite{LogisticMap} and later by Devaney \cite{DevaneyScience1987}, discrete dynamical systems (iterated maps) can also have overwhelmingly complex behavior in a simple compact form.  While PDEs efficiently describe aggregate behavior of many particles, maps can simulate single trajectories with very little computational expense.  This makes it computationally possible to model both single trajectories and aggregate behavior of combinations of many trajectories.  This is not a new concept, however; if we go back to our diffusion equation example, Einstein \cite{EinsteinDiffusion} developed a theory of Brownian motion by modeling the trajectories of individual particles and the probabilistic nature in which the direction changes.  Later Geisel and others \cite{DeterministicDiffusionReview} discovered diffusion could be induced through deterministic chaotic processes rather than just stochastic processes.

A single macro-scale particle exhibiting wave-like phenomena has been observed in walking droplet dynamics as shown by Harris \textit{et al.} \cite{HMFCB13} with several others delineated in the reviews by Bush and collaborators \cite{Bush15a, Bush15b, BushOza20_ROPP}.  While these examples are intriguing, abstracting away from the fluid mechanical experiments is necessary due to the differences in hydrodynamic and quantum processes.  Works by Papatryfonos \textit{et al.} \cite{PRBNBL_Tunneling2022}, Durey and Bush \cite{DureyBush2021, DureyBush-FieldTheory2020}, Durey \textit{et al.} \cite{DTB20}, Dagan and Bush \cite{DaganBush-FieldTheory2020}, Valani \textit{et al.} \cite{Valani2021}, Valani \cite{Valani-LorenzPRE2022, Valani-LorenzChaos2022}, and Perks and Valani \cite{PerksValani-Lorenz2023}, have shown the efficacy in abstracting away from experiments.  In recent years there have also been ensemble theories of pilot-wave theory inspired by walking droplets \cite{DaganBush-FieldTheory2020, DaganEnsemble2023} and attractor-driven quantum-like behavior \cite{ValaniQuantum-like2024}, which share some similar approaches to the present study.

In this paper we abstract away even further from fluid mechanics to consider the dynamical behavior of single-particle deterministic trajectories that aggregate (or ``ensemble'') to quantum-like statistics in three canonical examples:  motion in 1-dimensional free space, double slit diffraction in 3-dimensions, and superluminal barrier traversal in 1-dimension.  Moreover, we show various dynamical regimes can be accessed through bifurcations by varying $K$, such as the classical limit.

Motion in free space for a single classical particle is fairly simple, but a quantum particle behaves in a far more complex manner.  The studies of Heisenberg and others have shown an inverse relation between the uncertainty in momentum and that of position \cite{HeisenbergUncertainty, KennardUncertainty, WeylUncertainty}.  Further, in classical mechanics, diffraction can only be experienced by a wave, such as in Young's double slit \cite{YoungDoubleSlit}.  We would not expect single classical particles to create fringe like patterns unless they were severely manipulated.  Even further still, superluminal barrier traversal \cite{WinfulTunneling2006} is very much in the realm of waves.  It cannot be reproduced through kinematic ``tunneling'', but similar effects can appear in non-quantum waves \cite{HupertOttElectromagneticTunneling1966}.  While these examples do not prove that a discrete dynamical model of single particle trajectories is how nature behaves at the quantum scale, it shows that a wave function collapse is perhaps not necessary to interpret quantum mechanics.

%---%

%---MODELING---%
\section{Model setup}\label{Sec: Model}

In the investigation of Rahman \cite{RahmanAnnulus2023} they develop a hydrodynamic-kinematic model of a bouncing droplet on an annular fluid bath above the Faraday wave instability.  The model shows close agreement with experiments, and importantly, provides us with the machinery to develop a discrete dynamical model exhibiting diffusive behavior.  At each interaction with the wavefield the particle motion is determined by the slope of the wavefield and the previous velocity of the droplet, which is what the differential equations between impact model \cite{RahmanAnnulus2023}.  However, this can be abstracted away by observing the likelihood of the various interactions (see Sec. \ref materials for the detailed derivation).  This could be accomplished with a stochastic process or a deterministic dynamical system with the desired distribution in iterate space.  Fortuitously, the latter was found to be a uniform distribution through a trial and error process yielding the model
\begin{subequations}
\begin{align}
\tau_{n+1} &= 4\tau_n(1 - \tau_n),\label{Eq: Chaotic map 1}\\
t_{n+1}  &= t_n + \tau_{n+1}\sqrt{2} \mod 1, \label{Eq: Chaotic map 2}\\
v_{n+1} &= C\left[v_n + K\cos(2\pi t_{n+1})\sin(\omega x_n)e^{-\nu |v_n|}\right], \label{Eq: v map}\\
x_{n+1} &= v_{n+1} + x_n; \label{Eq: x map}
\end{align}
\label{Eq: TheModel}
\end{subequations}
with $C = 1/3$, $K = 95/2\pi$, and $\nu = 1/2.1$ from \cite{Rahman18} to agree with both the hydrodynamic-kinematic model and experiments in \cite{RahmanAnnulus2023} (see Sec. \ref{Sec: Justification for (3)} for details).  The chaotic map \eqref{Eq: Chaotic map 2} with intermediate updates \eqref{Eq: Chaotic map 1} produce the uniformly distributed iterates on the unit interval.  These drive the velocity/momentum map \eqref{Eq: v map} and position map \eqref{Eq: x map}, which were inspired by the standard map \cite{StandardMap}.  We recall that the standard map is derived from a Poincar\'{e} section of the kicked rotator \cite{KickedRotator}.  Interestingly, the quantum kicked rotator has been experimentally realized in super-cooled atoms \cite{RaizenKickedRotator1, AuklandKickedRotator, RaizenKickedRotator2}.  In addition, the trajectories of \eqref{Eq: TheModel} are reminiscent of possible trajectories in Feynman's path integral interpretation of quantum mechanics  \cite{FeynmanPathIntegral}.

\begin{figure}[htbp]
\centering
\includegraphics[width = \textwidth]{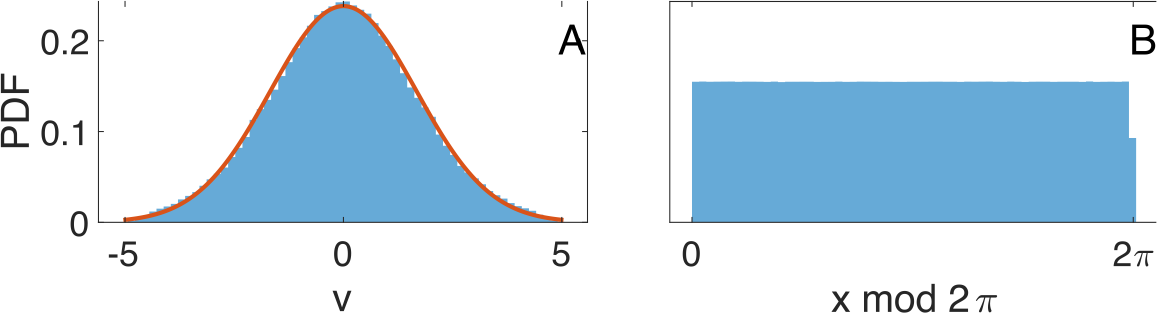}
\caption{A particle experiencing diffusive behavior in 1-dimension under the model \eqref{Eq: TheModel} with $K = 95/2\pi$ and initial conditions $t_0 = v_0 = 0$ and $\tau_0 = x_0 = 0.1$.  \textbf{(A)}  Velocity histogram of the particle converges to a Gaussian with a standard deviation of $\sigma_v \approx 1.675$  (see Sec. \ref{Sec: Properties of velocity distribution} for further statistical analysis).  \textbf{(B)}  Confining the particle to an annulus produces the expected distribution for a diffusive process.  (see \cite{Movie1} for the full simulation movie.)}\label{Fig: Diffusion}
\end{figure}

In Fig. \ref{Fig: Diffusion} we illustrate the movement of a single particle driven by \eqref{Eq: TheModel} on an annulus.  The particle changes velocity chaotically with a Gaussian distribution indicative of diffusive behavior.  Across the entire annulus, as expected for a diffusive particle, the particle visits every point in the domain with equal probability.
%
%%%---Figure of velocity and position histograms for a particle experiencing Brownian motion

%---RESULTS---%
\section{Three canonical examples}\label{Sec: Results}

Let us consider three examples of quantum effects that particle trajectories of \eqref{Eq: TheModel} exhibit: motion in free space, double slit diffraction, and superluminal barrier traversal.

\begin{figure}[htbp]
\centering
\includegraphics[width = \textwidth]{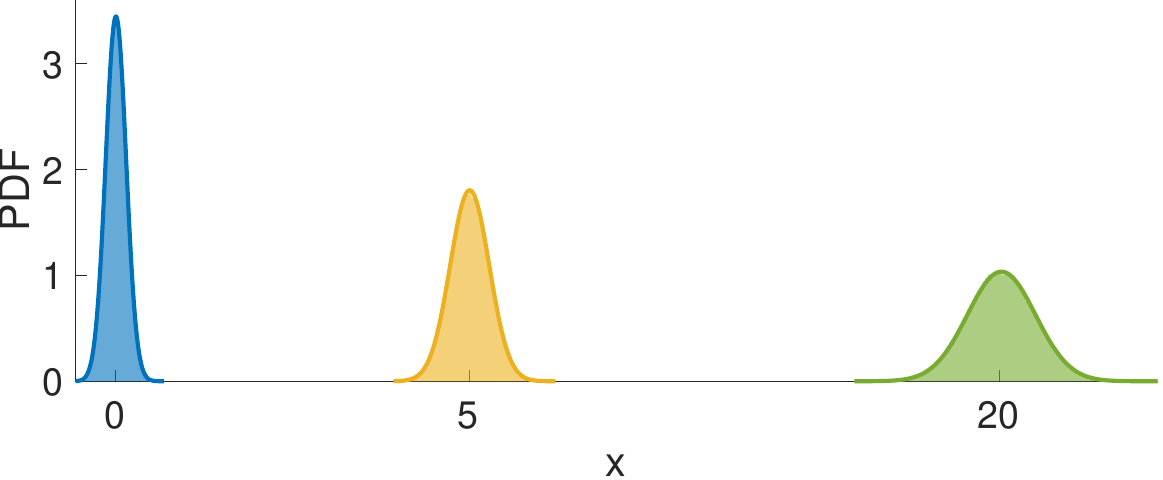}
\caption{Aggregate behavior of $10^5$ particles moving from left to right with a constant mean velocity of $\mu_v = 0.1$ and a perturbation governed by \eqref{Eq: TheModel} setting $\omega = 1$, $K = 10$, and initializing $t_0,\tau_0 \in (0,1)$ as a different random number for each individual particle, which induces a standard deviation of $\sigma_v \approx 0.032$.  As the group of particles move their distribution spreads out similar to $\hat{H} = \partial^2/\partial x^2$ in \eqref{Eq: Schrodinger}.  (see \cite{Movie2} for the full simulation movie.)}\label{Fig: FreeSpace}
\end{figure}

In Fig. \ref{Fig: Diffusion} we observed a particle experiencing diffusive behavior with a Normally distributed velocity.  In the following example we propel individual particles in 1-dimension from left to right with the same median velocity of $\mu_v = 0.1$ under the influence of \eqref{Eq: TheModel}.  We also set $K = 10$ for the remainder of the section for the sake of arithmetic simplicity.  We scale the velocity and substitute into \eqref{Eq: x map}
\begin{equation}
x_{n+1} =  x_n + (v_{n+1}/50 + \mu_v),
\label{Eq: Velocity Normalization}
\end{equation}
which induces a standard deviation of $\sigma_v \approx 0.032$.  As expected, in Fig. \ref{Fig: FreeSpace}, we observe the distribution of the particles spread as the group of particles move to the right.  At $x = 0$, $x = 5$, and $x = 20$, the standard deviations of the distributions are $\sigma_x \approx 0.118$, $\sigma_x \approx 0.223$, and $\sigma_x \approx 0.388$.  This is similar to what we might expect, qualitatively, from setting the Hamiltonian, $\hat{H} = \partial^2/\partial x^2$, in \eqref{Eq: Schrodinger}, and is reminiscent of the theoretical findings of Heisenberg \cite{HeisenbergUncertainty}, Kennard \cite{KennardUncertainty}, and Weyl \cite{WeylUncertainty}.

%%%---Figure of free space histograms for a with constant velocity

\begin{figure}[htbp]
\centering
\includegraphics[width = \textwidth]{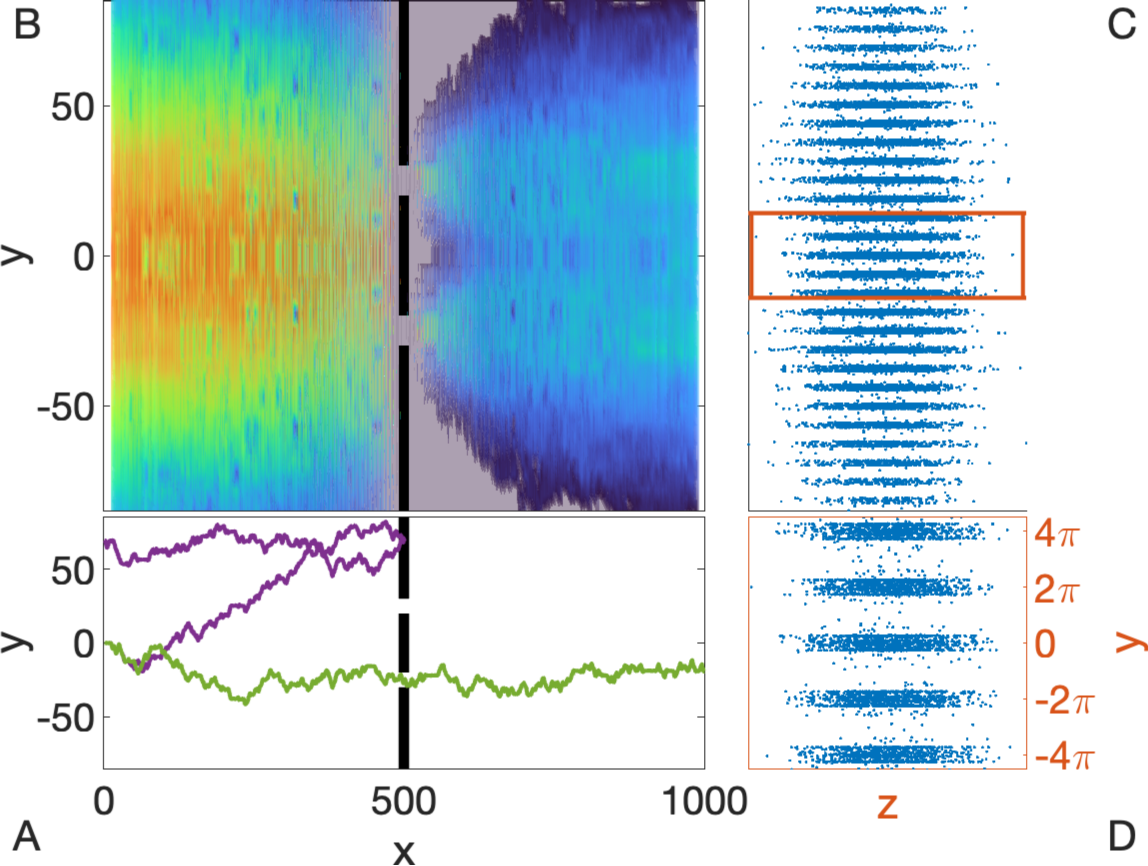}
\caption{Fringes forming \textbf{(C, D)} from the aggregation \textbf{(B)} of many individual trajectories \textbf{(A)}.  \textbf{(A)} Two representative individual particle trajectories from setting $\omega = 1$ and initializing $t_0,\tau_0 \in (0,1)$ as a different random number for each particle.  If a particle gets through one of the slits, we reinitialize $t_0 = 0, \tau_0 = 0.1$.  A screen is located at $x=1000$ with particles encountering a barrier with a double slit halfway at $x = 500$.  One trajectory goes through one of the slits and the other reflects back.  \textbf{(B)} An aggregate heat map of $10^4$ trajectories showing wave-like patterns similar to that of Philippidis \textit{et al.} \cite{PDH1978_Bohm}.  \textbf{(C)} Points from 3-dimensional trajectories (staring from an initial $10^4$ points) that have gone through the slits and intersected with the screen.  \textbf{(D)} Zoomed-in plot of \textbf{(C)} showing fringe-like patterns.}\label{Fig: DoubleSlit}
\end{figure}

In the next example we shoot individual test particles at a screen located at $x = 1000$ with a barrier at $x = 500$.  We assume the velocity in the $x$ direction is constant compared to that of the $y$ (horizontal) and $z$ (vertical) directions since the $x$ direction is the predominant direction of motion with a mean velocity several orders of magnitude larger than the variance in the other two directions.  We evolve the particle in the $y$ and $z$ directions according to \eqref{Eq: TheModel}. The barrier contains two slits with centers at $y = -25$ and $y = 25$ having a width of $\Delta y = 10$ each.  We set $\omega = 1$ in \eqref{Eq: TheModel}, and initialize $t_0, \tau_0 \in (0,1)$ as a different random number for each particle.  Some trajectories make it through one of the two slits and others hit the barrier and reflect back (Fig. \ref{Fig: DoubleSlit}(a)).  If a trajectory reaches the slits, we reinitialize $t_0 = 0, \tau_0 = 0.1$ for the motion in the $y$ direction (but not in the $z$ direction).  After many impacts with particles reaching the screen a pattern reminiscent of double-slit diffraction appears (Fig. \ref{Fig: DoubleSlit}(c, d)).  From an aggregation of both the reflected and penetrated trajectories we observe a wave-like pattern (Fig. \ref{Fig: DoubleSlit}(b)) including the quintessential triangle of low particle density between the two slits similar to the numerical experiments on Bohmian mechanics by Philippidis \textit{et al.} \cite{PDH1978_Bohm}.

%%%---Figure of double slit

It should be noted that the fringe pattern is predicated on the reinitialization $t_0 = 0, \tau_0 = 0.1$ and not the slits themselves.  That is, unlike a plane wave $\Psi$ evolving under
\begin{equation}
\frac{\partial^2 \Psi}{\partial t^2} = c^2\del^2\Psi
\label{Eq: Wave}
\end{equation}
encountering a spatial barrier with two slits, the model \eqref{Eq: TheModel} necessitates a reintialization in order to create the fringe pattern, and if a reinitialization occurs, the fringes appear regardless of the presence of the slits.  Indeed the reinitialization can be interpreted as an update to the Hamiltonian, $\hat{H}$, of \eqref{Eq: Schrodinger}.  However, unlike \eqref{Eq: Schrodinger}, a systematic updating procedure is not immediately obvious.  What is obvious is that these deterministic trajectories produce the fringe pattern with the only update being the initial condition of the chaotic maps (\ref{Eq: Chaotic map 1}, \ref{Eq: Chaotic map 2}).

\begin{figure}[htbp]
\centering
\includegraphics[width = \textwidth]{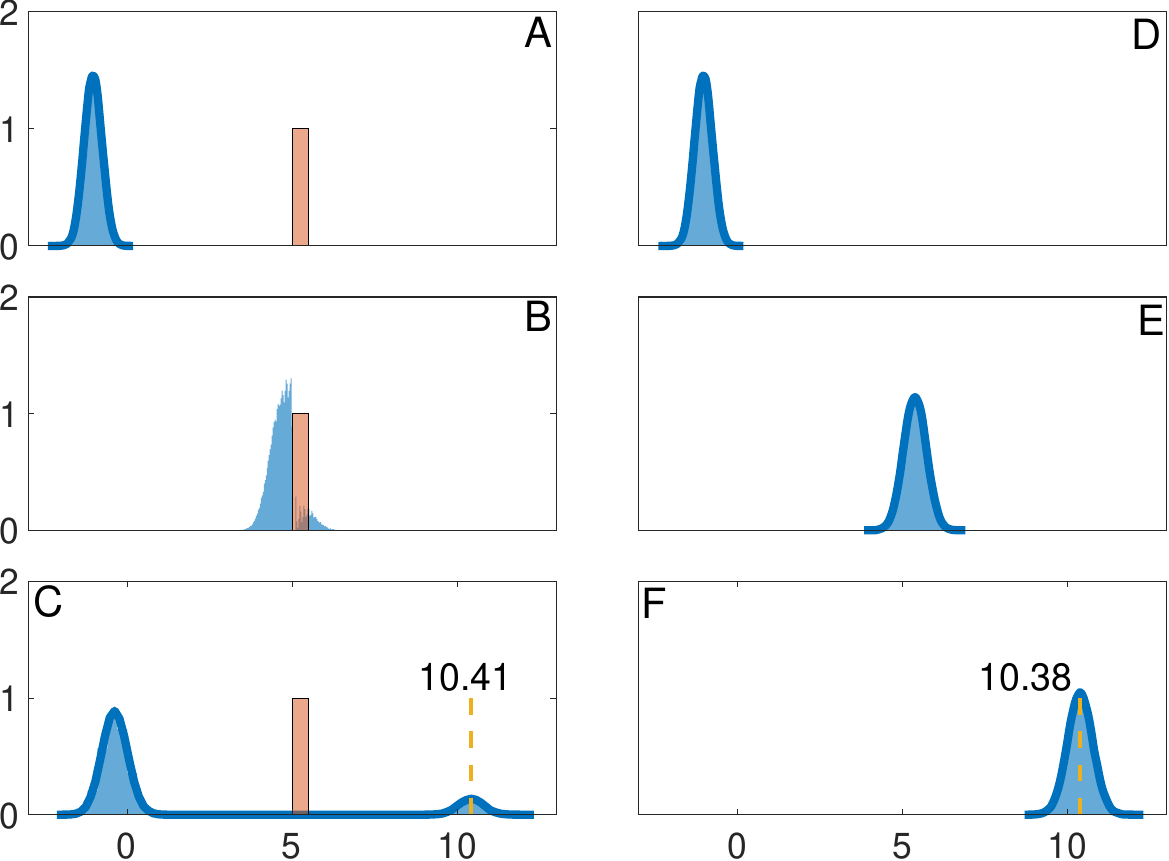}
\caption{Tunneling through an energy barrier \textbf{(A, B, C)} with associated unobstructed motion \textbf{(D, E, F)}.  \textbf{(A, D)} Aggregate behavior of $10^5$ particles moving to the right just as in Fig. \ref{Fig: FreeSpace}.  \textbf{(B, E)} Moments after the wave-packet reaches the barrier in \textbf{(A, B, C)} and the associated barrier free dynamics in \textbf{(D, E, F)}.  \textbf{(C, F)} Several iterations after colliding with the barrier the mean of the tunneled distribution in \textbf{(A, B, C)} is ahead of the mean of the unobstructed distribution in \textbf{(D, E, F)}.  (see \cite{Movie3} for the full simulation movie.)}\label{Fig: Tunneling}
\end{figure}

The final example is perhaps the most surprising.  We shoot individual particles, exactly as in the first example with the trajectory governed by \eqref{Eq: TheModel} with the scaled velocity \eqref{Eq: Velocity Normalization}, at a finite energy barrier, similar to setting $\hat{H} = \partial^2/\partial x^2 + V(x)$ in \eqref{Eq: Schrodinger} where $V(x)$ is the height of the barrier at position $x$.  The height of the barrier is equivalent to the mean kinetic energy of the particles with a width of $\Delta x = 0.5$ (Fig. \ref{Fig: Tunneling}(A)).  As expected most particles bounce back, however those with a kinetic energy greater than the barrier do indeed traverse through the barrier onto the other side.  When we observe the distribution of the particles over time, we notice a wave-like spike in the distribution near the barrier (Fig. \ref{Fig: Tunneling}(B)) before the distribution propagates back in the direction whence it came.  We also observe the mean of the distribution that went through the barrier (Fig. \ref{Fig: Tunneling}(C)) to be further along than the mean of the unimpeded particle (Fig. \ref{Fig: Tunneling}(F)), which is reminiscent of superluminal barrier traversal due to the Hartman effect \cite{HartmanTunneling1962}.

%%%---Figure of tunneling

While it is of no surprise that waves would experience the Hartman effect \cite{HartmanTunneling1962}, such as electromagnetic waves \cite{HupertOttElectromagneticTunneling1966}, single particle trajectories aggregating to wave-like behavior potentiates the possibility of obviating a wave function collapse to interpret the duality of an evolving probability wave of states and that of precise measurement.  The observed superluminal barrier traversal in Fig. \ref{Fig: Tunneling} promises to be a first step towards violating Bell's inequalities \cite{BellEPR1964, BellRMP1966} via statistical independence while preserving locality.

\section{Varying kick strength in a 1-D box}\label{Sec: Particle In A Box}

\begin{figure*}[htbp]
\centering
\includegraphics[width = \textwidth]{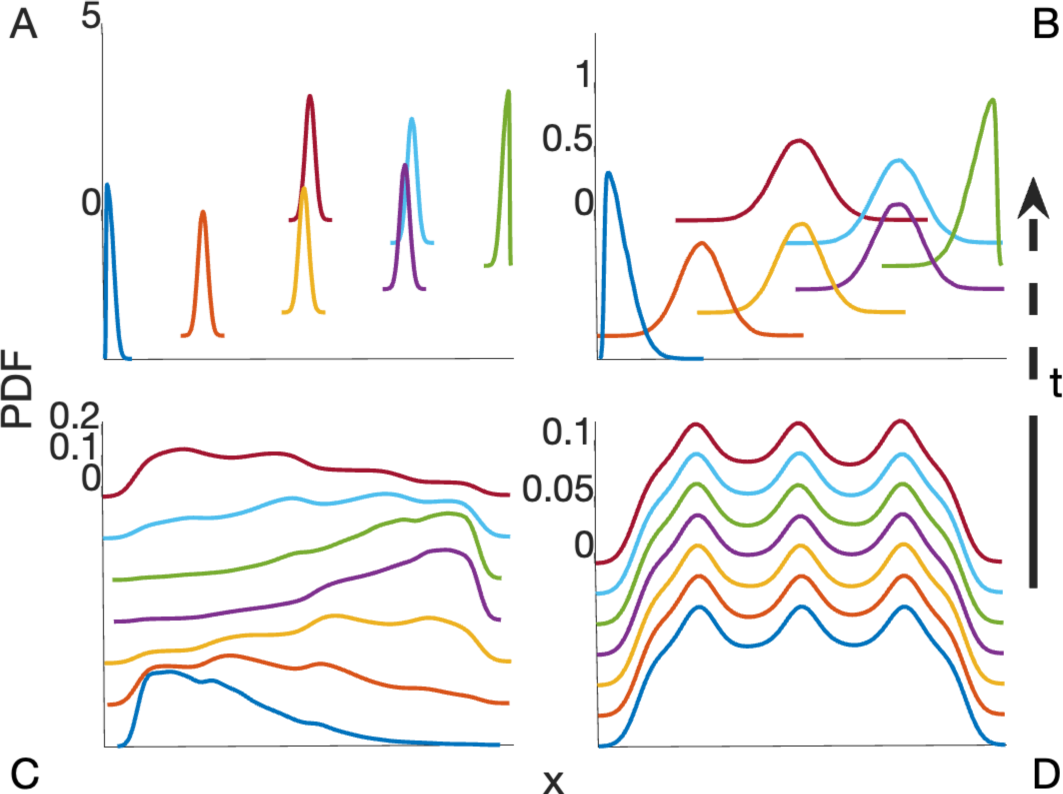}
\caption{Temporal evolution (in the increasing applicate direction into the page) of aggregate histograms (ordinate) of $10^5$ particles under the influence of \eqref{Eq: TheModel} with an initial rightward momentum in a 1-Dimensional box (abscissa) from $x = -5$ to $x = 5$ with reflective boundaries.  \textbf{(A)} For $K = 0$ the particle distribution behave as a classical object, such as a ball, by maintaining its shape unless acted on by an external potential (such as the reflective boundary).  \textbf{(B)} For $K = 10$ the histogram spreads out in space as the particles move with a constant mean speed (reminiscent of the uncertainty principle \cite{HeisenbergUncertainty}).  At the boundary the histogram interaction with the potential is also more wave-like.  \textbf{(C)}  For $K = 10^2$ a traveling wave is induced by the initial rightward momentum at first, and then by the potential at the reflective boundaries.  \textbf{(D)}  For $K = 10^3$ a standing wave with three peaks and two troughs is excited between the boundaries.    (see \cite{Movie4} for the full simulation movie.)}\label{Fig: ParticleInABox}
\end{figure*}

Similar to Feynman's path integral interpretation \cite{FeynmanPathIntegral}, the present formulation reaches a classical limit as $K \rightarrow 0$.  Moreover, as $K$ is increased, the system \eqref{Eq: TheModel} bifurcates to other dynamical regimes.  In Fig. \ref{Fig: ParticleInABox} we observe classical behavior, quantum-like behavior, traveling waves, and standing waves.  Similar to Sec. \ref{Sec: Results}, we propel $10^5$ individual Normally distributed particles with mean $\mu_x = 0$ and $\sigma_x \approx 0.119$ under the influence of \eqref{Eq: TheModel} with a positive initial speed of $v = 0.1$.

When $K = 0$, the distribution remains unchanged and maintains the initial standard deviation even after repetitive interactions with the reflective boundary.  The only instance when the distribution deviates from that of the initial histogram is during the bounces at the boundary.  During these interactions, the distribution gets squeezed similar to a ball bouncing on a wall; i.e., a distribution of classical particles.

When $K = 10$, the distribution spreads out just as in free space.  After five bounces the distribution spreads out to a standard deviation of $\sigma_x \approx 0.632$.  In addition, heuristically, the interaction with the boundary is more wave-like than that of the $K = 0$ case.  For the $K = 10^2$ and $K = 10^3$ cases we observe dynamics that are distinctly wave-like.  When $K = 10^2$ we observe traveling wave behavior, and when $K = 10^3$ we observe a distribution of particles akin to a standing wave.

%%%---Figure of particle in a box

% - CONCLUSION - %

\section{Discussions}\label{Sec: Conclusion}

Inspired by deterministic diffusion \cite{DeterministicDiffusionReview} and walking droplets \cite{CouderFort06}, in the present work we develop a deterministic model of particle trajectories \eqref{Eq: TheModel} driven by a chaotic map \eqref{Eq: Chaotic map 2} with a uniform distribution on the unit interval.  The model\eqref{Eq: TheModel} has a form similar to that of the standard map \cite{StandardMap}, which is derived from the kicked rotator \cite{KickedRotator}; a canonical example of quantum chaos.  The deterministic trajectories behave like systems of walking droplets, but also serve to abstract away from the fluid mechanics.  By using this abstraction we can aggregate over many trajectories and observe quantum-like behavior.  This is illustrated in three canonical examples: particle motion in free-space, double slit diffraction, and superluminal barrier traversal.

For the first example we propel single particles with a constant mean velocity and standard deviation induced by \eqref{Eq: Velocity Normalization}.  We then aggregate all of the trajectories and observe the distribution of particles spreading out as the mean of the distribution translates.  In our second example we shoot individual particles, according to \eqref{Eq: TheModel} in the $y$ and $z$ directions and assume a constant velocity at a much larger order of magnitude in the $x$ direction, at a screen and we place a barrier in the middle with two vertical slits.  If the particle goes through one of the slits we reinitialize the model \eqref{Eq: TheModel} in the $y$ direction (but not in the $z$ direction).  After aggregating over many trajectories a fringe pattern appears on the screen.  Although it requires a reinitalization of the chaotic maps (\ref{Eq: Chaotic map 1}, \ref{Eq: Chaotic map 2}), a reinitialization is also required for the Schr\"{o}dinger equation \eqref{Eq: Schrodinger} in the form of an updated Hamiltonian $\hat{H}$.  For the final example we shoot individual particles in the same way as the first example.  Most particles bounce back, but some go through the barrier due to the chaotically changing velocity.  Interestingly, the mean of the distribution of the tunneled particles is further along than if a barrier was not present, which is reminiscent of the Hartman effect \cite{HartmanTunneling1962}.  Finally, we vary $K$, which by proxy changes the energy of the system.  Increasing $K$ causes the system to bifurcate from classical behavior to wave-like behavior including increasing uncertainty in the position distribution, traveling waves, and standing waves.

In this paper we do not claim the present model is a complete description of nature at the quantum scale.  Rather, the examples provided herein show that aggregated deterministic particle trajectories driven by a simple chaotic model can produce wave-like behavior often thought of as exclusive to the quantum world.  If we can let go of statistical independence, then we can potentially develop a causal theory of quantum mechanics that preserves locality and violates Bell's inequality \cite{BellEPR1964}.  With the right model, the causal theory could provide a deterministic description of complex seemingly paradoxical phenomena such as the delayed choice quantum eraser \cite{DelayedChoiceEraser}.  Further, a causal theory could be the key to finally rectifying general relativity and quantum mechanics.  Moreover, as shown by Leghtas \textit{et al.} \cite{LeghtasQuantumDynSysSience2015}, a deeper understanding of connections between classical and quantum dynamical systems can contribute to improvements in quantum information techniques, such as error correction.

\section{Acknowledgments}
A.R. thanks Z. Nicolaou, G. Ferrandez-Quinto, and J. N. Kutz for fruitful discussions.  Thanks are also due to G. Ferrandez-Quinto and I. C. Christov for reading through the preliminary manuscript.  A.R. appreciates the support of the applied mathematics department and AI institute at University of Washington.

\bibliographystyle{unsrt}
\bibliography{nsBM}

\begin{thebibliography}{10}

\bibitem{BohmVigier}
D.~Bohm and J.P. Vigier.
\newblock Model of the causal interpretation of quantum theory in terms of a
  fluid with irregular fluctuations.
\newblock {\em Phys. Rev.}, 96:208, 1954.

\bibitem{AdvancesPilotWave}
J.~Croca P.~Castro, J. W. M.~Bush, editor.
\newblock {\em Advances in Pilot Wave Theory}.
\newblock Springer, 2024.

\bibitem{DonadiHossenfelder}
Sandro Donadi and Sabine Hossenfelder.
\newblock Toy model for local and deterministic wave-function collapse.
\newblock {\em Physical Review A}, 106:022212, 2022.

\bibitem{deBrogliePilotWave1927}
L.~de~Broglie.
\newblock La m\'{e}canique ondulatoire et la structure atomique de la
  mati\'{e}re et du rayonnement.
\newblock {\em J. Phys. Rad.}, 8:225--241, 1927.

\bibitem{HallDeterministic2010}
M.~J.~W. Hall.
\newblock Local deterministic model of singlet state correlations based on
  relaxing measurement independence.
\newblock {\em Phys. Rev. Lett.}, 105:250404, 2010.

\bibitem{Schrodinger}
E.~Schrodinger.
\newblock An undulatory theory of the mechanics of atoms and molecules.
\newblock {\em Phys. Rev.}, 28(6):1049--1070, 1926.

\bibitem{LogisticMap}
R.~M. May.
\newblock Simple mathematical models with very complicated dynamics.
\newblock {\em Nature}, 26(5560):457, 1976.

\bibitem{DevaneyScience1987}
R.~L. Devaney.
\newblock Chaotic bursts in nonlinear dynamical systems.
\newblock {\em Science}, 235(4786):342--345, 1987.

\bibitem{EinsteinDiffusion}
A~Einstein.
\newblock \"{U}ber die von der molekularkinetischen theorie der w\"{a}rme
  geforderte bewegung von in ruhenden fl\"{u}ssigkeiten suspendierten teilchen.
\newblock {\em Annalen Der Physik}, 17(8):549--60, 1905.

\bibitem{DeterministicDiffusionReview}
T.~Geisel.
\newblock Deterministic diffusion: A chaotic phenomenon.
\newblock {\em Europhys. News}, pages 5--8, 1984.

\bibitem{HMFCB13}
D.~M. Harris, J.~Moukhtar, E.~Fort, Y.~Couder, and J.~W.~M. Bush.
\newblock Wavelike statistics from pilot-wave dyanmics in a circular corral.
\newblock {\em Phys. Rev. E}, 88:011001(R), 2013.

\bibitem{Bush15a}
J.W.M. Bush.
\newblock Pilot-wave hydrodynamics.
\newblock {\em Ann. Rev. Fluid Mech.}, 49:269--292, 2015.

\bibitem{Bush15b}
J.W.M. Bush.
\newblock The new wave of pilot-wave theory.
\newblock {\em Physics Today}, 68(8):47--53, 2015.

\bibitem{BushOza20_ROPP}
J.~W.~M. Bush and A.~U. Oza.
\newblock Hydrodynamic quantum analogs.
\newblock {\em Reports on Progress in Physics}, 84:017001, 2021.

\bibitem{PRBNBL_Tunneling2022}
K.~Papatryfonos, M.~Ruelle, C.~Bourdiol, A.~Nachbin, J.~W.~M. Bush, and
  M.~Labousse.
\newblock Hydrodynamic superradiance in wave-mediated cooperative tunneling.
\newblock {\em Commun. Phys.}, 5:142, 2022.

\bibitem{DureyBush2021}
M.~Durey and J.~W.~M. Bush.
\newblock Classical pilot-wave dynmaics: The free particle.
\newblock {\em Chaos}, 31:033136, 2021.

\bibitem{DureyBush-FieldTheory2020}
M.~Durey and J.~W.~M. Bush.
\newblock Hydrodynamic quantum field theory: The onset of particle motion and
  the form of the pilot wave.
\newblock {\em Front. Phys.}, 8:300, 2020.

\bibitem{DTB20}
M.~Durey, S.~E. Turton, and J.~W.~M. Bush.
\newblock Speed oscillations in classical pilot-wave dynamics.
\newblock {\em Proc. Roy. Soc. A}, 476(2239):1--23, 2020.

\bibitem{DaganBush-FieldTheory2020}
Y.~Dagan and J.~W.~M. Bush.
\newblock Hydrodynamic quantum field theory: the free particle.
\newblock {\em Comptes. Rendus. M\'{e}canique}, 348(6-7):555--571, 2020.

\bibitem{Valani2021}
R.~N. Valani, A.~C. Slim, D.~M. Paganin, T.~P. Simula, and T~Vo.
\newblock Unsteady dynamics of a classical particle-wave entity.
\newblock {\em Phys. Rev. E}, 104:015106, 2021.

\bibitem{Valani-LorenzPRE2022}
R.~N. Valani.
\newblock Anomalous transport of a classical wave-particle entity in a tilted
  potential.
\newblock {\em Phys. Rev. E}, 105:L012101, 2022.

\bibitem{Valani-LorenzChaos2022}
R.~N. Valani.
\newblock Lorenz-like systems emerging from an integro-differential trajectory
  equation of a one-dimensional wave--particle entity.
\newblock {\em Chaos}, 32(2):023129, 2022.

\bibitem{PerksValani-Lorenz2023}
J.~Perks and R.~N. Valani.
\newblock Dynamics, interference effects, and multistability in a lorenz-like
  system of a classical wave--particle entity in a periodic potential.
\newblock {\em Chaos}, 33(3):033147, 2023.

\bibitem{DaganEnsemble2023}
Y.~Dagan.
\newblock {\em Advances in Pilot Wave Theory}, volume 344 of {\em Boston
  Studies in the Philosophy and History of Science}, chapter Hydrodynamically
  Inspired Pilot-Wave Theory: An Ensemble Interpretation.
\newblock Springer, 2024.

\bibitem{ValaniQuantum-like2024}
R.~N. Valani and \'{A}.~G. L\'{o}pez.
\newblock Quantum-like behavior of an active particle in a double-well
  potential.
\newblock {\em arxiv:2401.05616}, 2024.

\bibitem{HeisenbergUncertainty}
W.~Heisenberg.
\newblock \"{U}ber den anschaulichen inhalt der quantentheoretischen kinematik
  und mechanik.
\newblock {\em Zeitschrift f\"{u}r Physik}, 43:172--198, 1927.

\bibitem{KennardUncertainty}
E.~H. Kennard.
\newblock Zur quantenmechanik einfacher bewegungstypen.
\newblock {\em Zeitschrift f\"{u}r Physik}, 44(4-5):326--352, 1927.

\bibitem{WeylUncertainty}
H.~Weyl.
\newblock Gruppentheorie und quantenmechanik.
\newblock {\em Monatshefte f\"{u}r Mathematik und Physik}, 36:A48--A52, 1929.

\bibitem{YoungDoubleSlit}
Thomas Young.
\newblock The bakerian lecture. experiments and calculation relative to
  physical optics.
\newblock {\em Philosophical Transactions of the Royal Society of London},
  94:1--16, 1804.

\bibitem{WinfulTunneling2006}
H.~G. Winful.
\newblock Tunneling time, the hartman effect, and superluminality: A proposed
  resolution of an old paradox.
\newblock {\em Phys. Rep.}, 436(1-2):1--69, 2006.

\bibitem{HupertOttElectromagneticTunneling1966}
J.~J. Hupert and G.~Ott.
\newblock Electromagnetic analog of the quantum-mechanical tunnel effect.
\newblock {\em Am. J. Phys.}, 34:260--265, 1966.

\bibitem{RahmanAnnulus2023}
Aminur Rahman.
\newblock Damped-driven system of bouncing droplets leading to deterministic
  diffusive behavior.
\newblock {\em Phys. Rev. E}, 108:035103, 2023.

\bibitem{Rahman18}
Aminur Rahman.
\newblock Standard map-like models for single and multiple walkers in an
  annular cavity.
\newblock {\em Chaos}, 28:096102, 2018.

\bibitem{StandardMap}
B.~V. Chirikov.
\newblock A universal instability of many-dimensional oscillator systems.
\newblock {\em Phys. Rep.}, 52:263, 1979.

\bibitem{KickedRotator}
B.~V. Chirikov, G.~Casati, F.~M. Izrailev, and J.~Ford.
\newblock {\em Lecture Notes in Physics}, chapter Stochastic behavior of a
  quantum pendulum under a periodic perturbation.
\newblock Springer-Berlin, 1978.

\bibitem{RaizenKickedRotator1}
F.~L. Moore, J.~C. Robinson, C.~F. Bharucha, Bala Sundaram, and M.~G. Raizen.
\newblock Atom optics realization of the quantum $\delta$-kicked rotor.
\newblock {\em Phys. Rev. Lett.}, 75(25):4598--4601, 1995.

\bibitem{AuklandKickedRotator}
H.~Ammann, R.~Gray, I.~Shvarchuck, and N.~Christensen.
\newblock Quantum delta-kicked rotor: Experimental observation of decoherence.
\newblock {\em Phys. Rev. Lett.}, 80(19):4111--4115, 1998.

\bibitem{RaizenKickedRotator2}
B.~G. Klappauf, W.~H. Oskay, D.~A. Steck, and M.~G. Raizen.
\newblock Observation of noise and dissipation effects on dynamical
  localization.
\newblock {\em Phys. Rev. Lett.}, 81(6):1203--1206, 1998.

\bibitem{FeynmanPathIntegral}
R.~P. Feynman.
\newblock Space-time approach to non-relativistic quantum mechanics.
\newblock {\em Rev. Mod. Phys.}, 20(2):367--387, 1948.

\bibitem{Movie1}
See supplemental material at [url will be inserted by publisher] for [movie 1:
  Simulation of a diffusive particle corresponding to fig. 1.].

\bibitem{Movie2}
See supplemental material at [url will be inserted by publisher] for [movie 2:
  Simulation of an aggregate of $10^5$ trajectories in free space corresponding
  to fig. 2.].

\bibitem{PDH1978_Bohm}
C.~Philippidis, C.~Dewdney, and B.~J. Hiley.
\newblock Quantum interference and th quantum potential.
\newblock {\em Il Nuovo Cimento della Societ\`{a} Italiana di Fisica},
  52(1):15--28, 1978.

\bibitem{HartmanTunneling1962}
T.~E. Hartman.
\newblock Tunneling of a wave packet.
\newblock {\em J. Appl. Phys.}, 33:3427--3433, 1962.

\bibitem{Movie3}
See supplemental material at [url will be inserted by publisher] for [movie 3:
  Simulation of an aggregate of $10^5$ trajectories tunneling through a
  potential barrier corresponding to fig. 4.].

\bibitem{BellEPR1964}
J.~S. Bell.
\newblock On the einstein podolsky rosen paradox.
\newblock {\em Physics}, 1(3):195--200, 1964.

\bibitem{BellRMP1966}
J.~S. Bell.
\newblock On the problem of hidden variables in quantum mechanics.
\newblock {\em Rev. Mod. Phys.}, 38:447, 1966.

\bibitem{Movie4}
See supplemental material at [url will be inserted by publisher] for [movie 4:
  Simulation of an aggregate of $10^5$ particle in a box trajectories for $k =
  0$, $k = 10$, $k = 10^2$, and $k = 10^3$ corresponding to fig. 5.].

\bibitem{CouderFort06}
Y.~Couder and E.~Fort.
\newblock Single-particle diffraction and interference at a macroscopic scale.
\newblock {\em Phys. Rev. Lett.}, 97:154101, 2006.

\bibitem{DelayedChoiceEraser}
Yoon-Ho Kim, R.~Yu, S.~P. Kulik, Y.~Shih, and M.~O. Scully.
\newblock A delayed "choice" quantum eraser.
\newblock {\em Phys. Rev. Lett.}, 84(1):1--5, 2000.

\bibitem{LeghtasQuantumDynSysSience2015}
Z.~Leghtas, S.~Touzard, I.~M. Pop, A.~Kou, B.~Vlastakis, A.~Petrenko, K.~M.
  Sliwa, A.~Narla, S.~Shankar, M.~J. Hatridge, M.~Reagor, L.~Frunzio, R.~J.
  Schoelkopf, M.~Mirrahimi, and M.~H. Devoret.
\newblock Confining the state of light to a quantum manifold by engineered
  two-photon loss.
\newblock {\em Science}, 347(6224):853--857, 2015.

\end{thebibliography}

\appendix

\section{Justification for argument in eq. (3)}\label{Sec: Justification for (3)}

\begin{figure}[htbp]
\centering
\includegraphics[width = 0.45\textwidth]{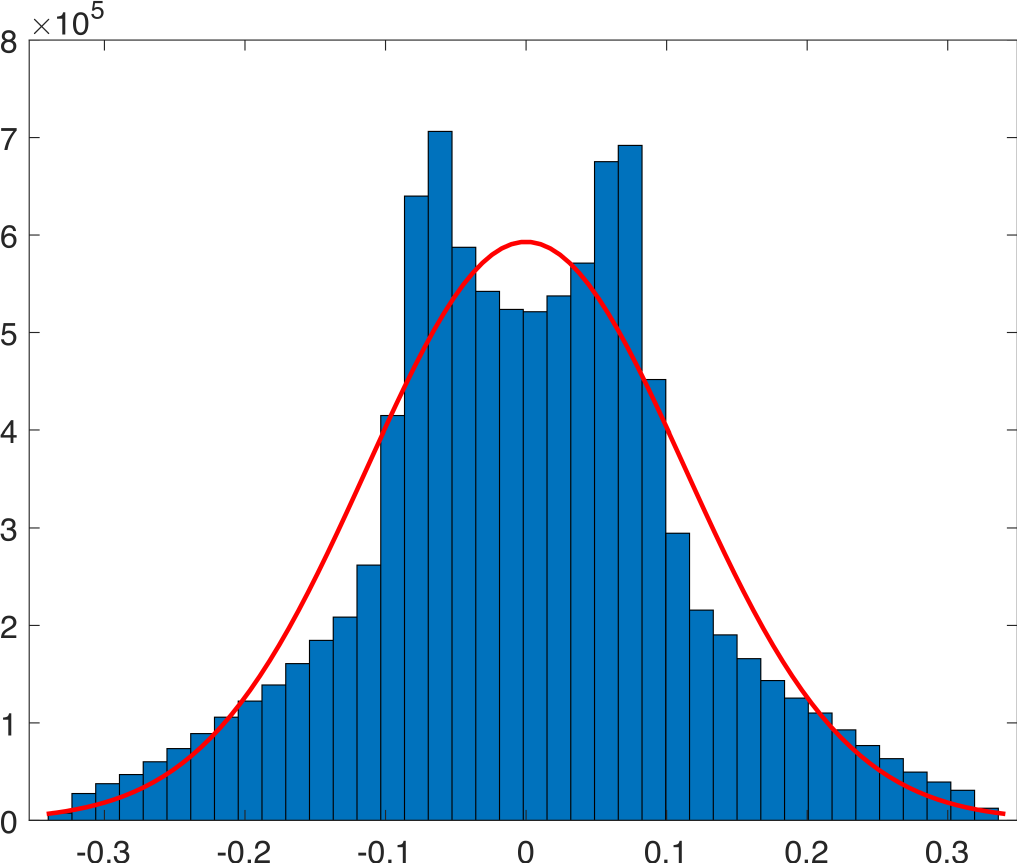}
\caption{Velocity distribution for the Standard map-like model in \cite{Rahman18}.}
\label{Fig: Standard map histogram}
\end{figure}

The Standard map inspired model of Rahman \cite{Rahman18}, while mathematically tractable and amenable to analysis, fails to agree with experiments of walking droplets on an annulus since the velocity distribution of the model in \cite{Rahman18} does not converge to a Gaussian as shown in Fig. \ref{Fig: Standard map histogram}.
However, it is still promising since the histogram is symmetric.  Therefore, the qualitative behavior of the model in \cite{Rahman18} is similar to that of the experiments in \cite{RahmanAnnulus2023}.

\begin{figure}[htbp]
\centering
\includegraphics[width = 0.45\textwidth]{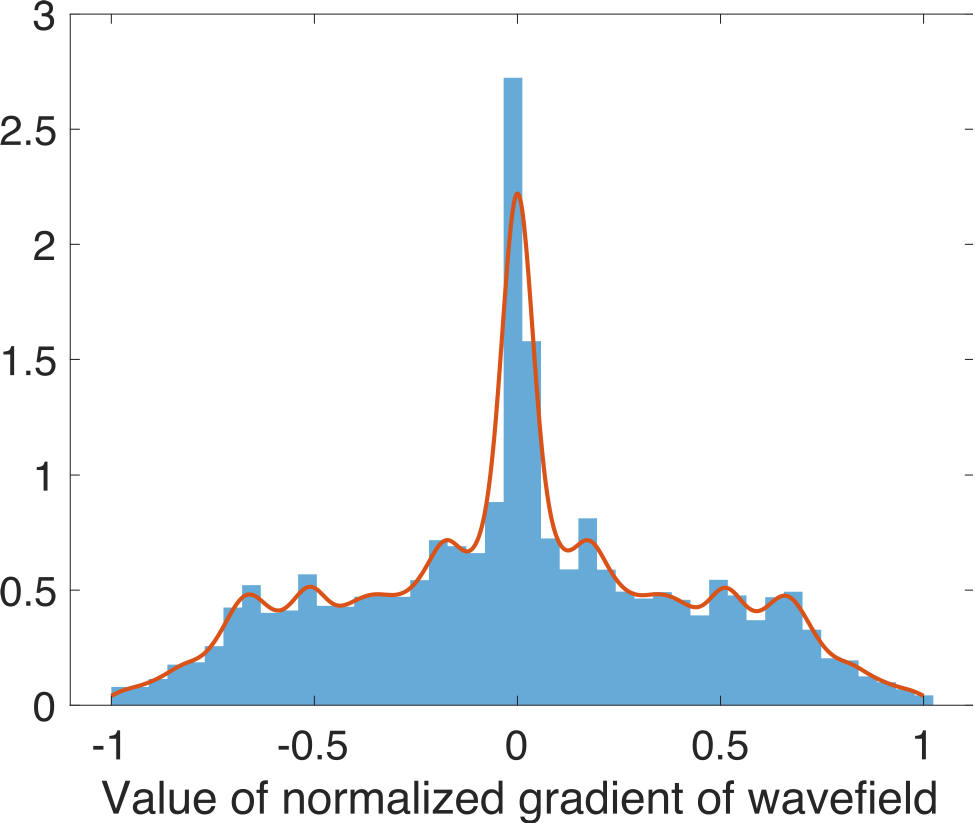}\qquad
\includegraphics[width = 0.45\textwidth]{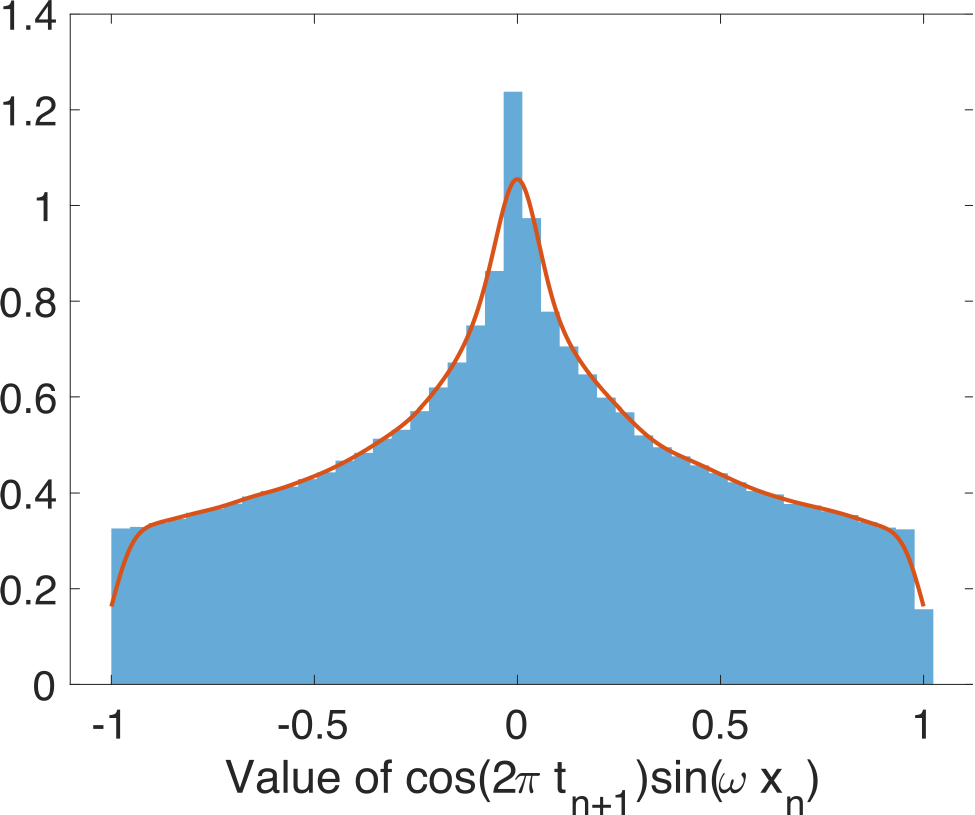}
\caption{Histograms \textbf{(top)} of the normalized wavefield gradient eq. (6) in [32] and \textbf{(bottom)} of $\cos(2\pi t_{n+1})\sin(\omega x_n)$.}
\label{Fig: wavefield gradient histograms}
\end{figure}

We hypothesize that the quantitative behavior can achieve closer agreement by replacing the $\sin\omega v_n$ term in \cite{Rahman18} with a function closer to that of the wavefield gradient, and more specifically the distribution of values of the sinusoidal terms at each iteration, $n$, should be similar to normalized wavefield gradient from experiments and the hydrodynamic-kinematic model eq. (6) in \cite{RahmanAnnulus2023}.  We observe that the two-step chaotic map from eq. (3), yields a similar distribution of values to that of eq. (6) in \cite{RahmanAnnulus2023} as shown in Fig. \ref{Fig: wavefield gradient histograms}.
More importantly, as shown in the main text of the manuscript, the histogram of velocity is a Gaussian similar to that of \cite{RahmanAnnulus2023}.

\section{Properties of the distributions in Fig. 1}\label{Sec: Properties of velocity distribution}

\begin{figure}[htbp]
\centering
\includegraphics[width = 0.45\textwidth]{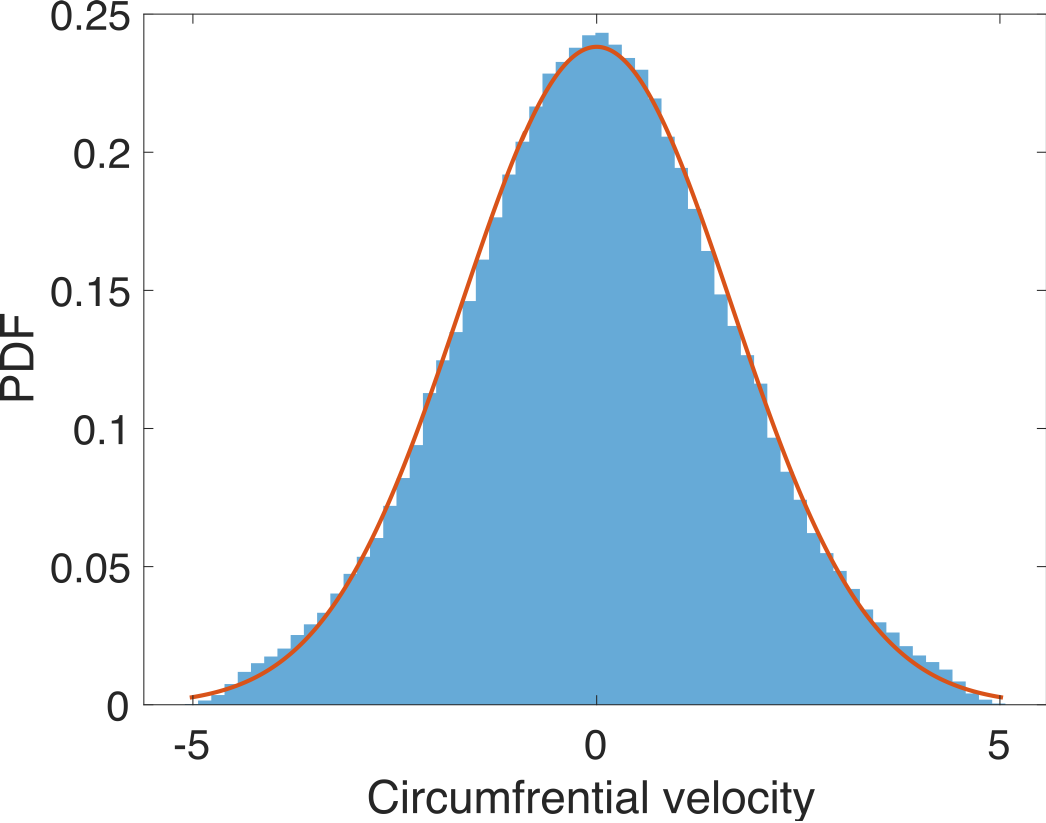}
\caption{Velocity distribution of eq. (3) indicating deterministic diffusion.}
\label{Fig: Velocity distribution}
\end{figure}

We consider the model eq. (3) in an annulus, which produces deterministic diffusive behavior.  When we take the histogram (Fig. \ref{Fig: Velocity distribution}) of the velocity distribution we get a mean of $\mu_v \approx 1.015\times 10^{-4}$ and a standard deviation of $\sigma_v \approx 1.675$.

\end{document}